# Basic Research, Lethal Effects: Military AI Research Funding as Enlistment

**David Gray Widder**

Digital Life Initiative, Cornell University

**Sireesh Gururaja**

School of Computer Science, Carnegie Mellon University

**Lucy Suchman**

Department of Sociology, Lancaster University

## Abstract

In the context of unprecedented U.S. Department of Defense (DoD) budgets, this paper examines the recent history of DoD funding for academic research in algorithmically based warfighting. We draw from a corpus of DoD grant solicitations from 2007 to 2023, focusing on those addressed to researchers in the field of artificial intelligence (AI). Considering the implications of DoD funding for academic research, the paper proceeds through three analytic sections. In the first, we offer a critical examination of the distinction between basic and applied research, showing how funding calls framed as basic research nonetheless enlist researchers in a war fighting agenda. In the second, we offer a diachronic analysis of the corpus, showing how a 'one small problem' caveat, in which affirmation of progress in military technologies is qualified by acknowledgement of outstanding problems, becomes justification for additional investments in research. We close with an analysis of DoD aspirations based on a subset of Defense Advanced Research Projects Agency (DARPA) grant solicitations for the use of AI in battlefield applications. Taken together, we argue that grant solicitations work as a vehicle for the mutual enlistment of DoD funding agencies and the academic AI research community in setting research agendas. The trope of basic research in this context offers shelter from significant moral questions that military applications of one's research would raise, by obscuring the connections that implicate researchers in U.S. militarism.

## Introduction

The United States Department of Defense (DoD) was arguably the first major investor in the field of artificial intelligence (AI), and remains one even amidst the ascendance of corporate players. Defense spending on AI nearly tripled from 2022 to 2023 (Henshall 2024), and in 2024 the DoD requested $1.8 billion for AI research, development, testing and evaluation.[1] As corporate capture of AI research receives increasing scrutiny,[2] the close association between academic AI research and the DoD has endured. What is funded affects what is created, but also the norms of the wider AI research ecosystem: in one example, research on the incentives governing academic Natural Language Processing (NLP) research shows how the DoD's switch from "patient" funding approaches to a benchmark culture meaningfully affected the course and culture of the NLP field (Gururaja et al. 2023).

In this paper, we explore a corpus of DoD grant solicitations issued between 2007 and 2023 (Bateyko & Levy, 2024), aided by NLP and information retrieval tools, selecting for calls that reference Artificial Intelligence and associated technologies. Focusing on the implications of DoD funding for academic research, we read these grant solicitations as a vehicle for the mutual enlistment of academic researchers and DoD research agencies in each others' agendas.[3] Broad Agency Announcements and other funding calls outline joint projects that align interests in the research community with military priorities, at the same time that those priorities are shaped by the promissory technologies on offer from academic and commercial research laboratories.

In what follows we set out our discussion in four parts. We begin with a description of the corpus and our methodological experiments with NLP-enabled search as a resource for interpretive analysis. We then proceed through three analytic sections. In the first, we offer a critical examination of the distinction between "basic" and "applied" research, showing how funding calls targeted towards the former, while they may make it easier for researchers to avoid thinking of their work as being directed towards lethal ends, nonetheless enlist researchers in a military agenda. In the second, we offer a diachronic analysis of the corpus, taking three series of Broad Agency Announcements calling for basic research as cases. Reading across these funding programs, we track a recurring formulation that we call the "one small problem" caveat. This is a rhetorical move in which affirmation of progress in military technologies is qualified by acknowledgement of outstanding problems, which become justification for additional investments in research. The third analytic section offers an analysis based on a subset of Defense Advanced Research Projects Agency (DARPA) grant solicitations making explicit reference to the use of AI in battlefield applications, attending to the agency's problem framings and imagined technological solutions.

---

[1] United States Department of Defense Fiscal year 2024 Budget Request
https://comptroller.defense.gov/Portals/45/Documents/defbudget/FY2024/FY2024_Budget_Request.pdf

[2] For analyses of the expansion of DoD funding for commercial research, particularly through Silicon Valley venture capital and major corporations in the tech sector, see Whittaker 2021, González 2024.

[3] We choose the term 'enlistment' for its connotations of military recruiting, though as we go on to discuss, the relation between the military and the research community is a reciprocal one. We take enlistment in this context as a synonym for 'enrolment' as that term has been developed within the science and technology studies (STS) literature. Callon (1984) cites enrolment as one among a set of processes through which problems are defined, solutions put forward, and actors recruited into positions and relations that work to sustain a particular definition of their collective situation.

Taken together, our analyses demonstrate the mutual enlistment of DoD funding agencies and the academic and commercial AI research community in setting research agendas. The DoD strategically mobilizes the category of "basic research" to reflect the interests and directions of the research community, which in turn inform the problem formulations and imagined solutions of DoD battlefield applications. Our analysis of grant solicitations over time, and with respect to their expressed desiderata for applications of AI in warfighting, shows how rather than describing two separate spheres, the distinction between basic and applied research works as a flexible tether that loosely, but consequentially, binds together the military and academic communities. The DoD's sociotechnical imaginaries and military objectives shape the directions and accountabilities of academic researchers as they pursue DoD-funded research projects. It follows that academic researchers funded by the DoD must engage with the ways in which they are implicated in the project of US military hegemony.

## Interpretive methods with models

This paper offers an interpretive analysis supported by computational methods of a corpus of grant solicitations issued between late 2007 and late 2023 from the US Department of Defense. Solicitations are calls for academics to submit research proposals for possible funding, thus aspirationally describing research a DoD agency hopes to fund.[4] Our original corpus comprised 46,175 solicitations automatically downloaded in bulk from the publicly available US government grants.gov database, which we filtered for those originating from the Department of Defense, narrowing our corpus down to 9292. We retained only documents in PDF format, discarding other formats, most but not all containing budget, worksheets, and other details,[5] creating a final dataset of 7187 documents.

We then made use of three computational methods to help examine the 7187 documents: (1) a searchable database of the dataset content, (2) a clustering method to identify temporal sequences of releases of the same solicitation, and (3) a topic model to help us characterize the types of language that appear in these documents. In identifying documents for closer reading, our aim was not a quantitative, distributional analysis but rather an examination of the aspirations articulated by the solicitations, as informed by prevalent directions in academic research in computer science. Our analyses are intended to be indicative, opening lines of investigation rather than offering a comprehensive survey of the corpus.

To elaborate briefly on these computational methods, we begin with tools for search. As even simple keyword searches of that many PDF documents quickly overwhelmed our personal computers, we uploaded the documents to an ElasticSearch database hosted on a central server. This allowed us to search the documents based on keywords or metadata, and to filter results based on dates or the DoD agency they originate from.

---

[4] The research presented here could clearly be deepened through an affiliated investigation of research contracts granted. DoD procurement records, as documents of the military-commercial complex devoted to surveillance and weaponry, are tracked in the work of Jack Poulson's Tech Inquiry.

[5] Microsoft Word documents were the riskiest to omit, given that they sometimes contained grant content. However, we found that they were mostly instructions or templates, and otherwise were often redundant to content in PDF documents.

Secondly, to explore how the dataset is structured temporally, we clustered[6] titles of solicitations into groups that differ only by a few characters – meaning that often clusters reflected solicitations whose titles only varied in their year of issue. This turned up recurring funding programs within the DoD grants ecosystem that are re-issued on a periodic basis. We focus on three such programs, detailed in a later section.

Thirdly, we created a topic model of these documents: a "statistical model[] for learning the latent structure in document collections" (Boyd-Graber et al., 2014), which we used to sort document snippets into thematic categories.[7] Even while this categorizes text with some degree of automation, many in the NLP community recognize the role of interpretive judgment in both building and making sense of their output. Given this, we approach topic modeling not as mechanically objective but as an exercise in trained judgment (Daston and Galison 2007). In particular, these judgments include constructing our model using bigrams like "artificial intelligence" or "battlefield surveillance," using 40 topics that appeared to be at the proverbial "elbow optimum"[8] and provided a good tradeoff between overly broad topics and topics that were too specific to represent recognizable similarities. . We offer examples of the kind of qualitative analysis these tools made possible below.

## "Basic Research" is morally intertwined with the battlefield

There are a variety of morally relevant distinctions that researchers draw when choosing whether to accept military funding, and if so, what particular projects to work on. Chief among them is the basic/applied distinction, where basic research is considered to be conducted without ostensible intentions for particular end uses, or more specifically in the case of DoD funding not connected to weapons or war as applied research might be (Kuipers 2021). The DoD's "Basic Research Office" website quotes Vannevar Bush's notorious report "Science: the Endless Frontier" to say that basic research "creates the fund from which the practical applications of knowledge must be drawn" (Department of Defense, n.d.). Our dataset provides examples of how the imagined distinction between basic and applied is maintained and elaborated by the DoD, and many more examples that suggest that the distinction's moral relevance is weaker than imagined. However questionable, the distinction has real consequences for the enlistment of researchers into DoD projects. It enrolls them into thinking like the DOD in their conceptualization of research problems and the applications they might be used for, in order to improve their chances of a successful grant proposal. It also provides "moral wiggle room" or plausible deniability, by providing ways of thinking about their work as basic and thus unconnected to end use (Widder and Nafus 2023), in this case for war, "battlefield management," or killing.

---

[6] We use the DBSCAN clustering algorithm, using the Levenshtein distance as the clustering metric. The Levenshtein distance, also referred to as "edit distance" refers to the number of operations in order to transform one string into another.

[7] Specifically, we fit a Latent Dirichlet Allocation (LDA) model to our corpus, in which documents are represented by a probability distribution over topics that are most likely represented within them, and each topic is represented by a probability distribution over words, with words more closely associated with that topic having a higher probability. The topic model is "trained," a process by which topics are inferred from the corpus, such that words selected from that topic could reproduce a document with the same distribution of words as a document labeled to be of that topic, on average.

[8] The "elbow optimum" refers to the point where increasing the number of topics no longer greatly improves the degree to which the topic model accurately represents the statistical distribution of the data. For a more thorough discussion of practical issues in clustering, see section 14.3.11 of Hastie, Tibishrani, and Friedman (2009).

## Context for the Basic / Applied distinction

The Congressional Research Service shows how DoD science and technology funding is broken into eight "budget activity" categories (Sargent, Jr, 2022). Categories 6.1 and 6.2 are "Basic" and "Applied" research respectively. Numbers from 2022 show that DoD funding makes up 41.2% of the entire federal budget for research and development, and within this DoD portion, Basic and Applied research together make up less than 10% of the expenditure, with the lion's share going to "operational systems development" and "advanced component development and prototypes" (Sargent, Jr, 2022). And yet within the overall small fraction, "basic" research is funded at about one third the level of "applied" research. Even if basic research is to be considered in some way morally exempt, in other words, it is at best a tiny adornment on a much larger arsenal of R&D funding directed more explicitly towards military ends.

University policies attempt to support free exchange of ideas while accepting military funding. As one example, Carnegie Mellon University (CMU) derives 42% of its research funding from the DoD (Carnegie Mellon University, 2023), accruing the third most DoD funding of any university (Folts, 2024). The university has a policy that by default this funding must be "unrestricted" in terms of publication, though it allows explicitly restricted and military-funded research to occur in its "nonacademic" "semi-autonomous units", and even in academic units by exception (Carnegie Mellon University, 1988). Debate within CMU's School of Computer Science about the ethics of such funding affirms an apparently morally relevant distinction between "weapons" and "non-weapons R&D", but capaciously defines the latter to include the development of the autonomous tactical vehicle named "Crusher", because no literal weapons were built onto it by the university (Carnegie Mellon University, n.d.) CMU also hosted a listening session for the DOD's Defense Innovation Board, which affirmed the DoD's goal to use AI to "fight and win in future wars" (Defense Innovation Board, 2019). Recommendations from the Defense Innovation Board included strengthening private sector and university connections, held up top tech companies and Elon Musk (by name) as inspiration, and recommended a "focus on exploratory research". In short, while basic/applied distinctions are seen as core to the university's mission and to the maintenance of academic ethics, our dataset allows us to examine how the basic/applied distinction works in DoD funding calls and the questions that this raises.

## How basic research is directed to military ends

Even when a program is framed as funding for basic research, there are indications that positioning a proposal in relation to DoD objectives will increase its chances of success. For example, the 2024 Vannevar Bush Faculty Fellowship call (#1)[9] says, in bold text, that the fellowship is intended for "ambitious 'blue sky' research", and opens by quoting the DoD Definition of "basic research", as "systematic study directed toward greater knowledge", "without specific applications […] or products in mind." This is immediately followed, however, by the qualification that such work must be "related to long-term national security needs." The stated objectives of the program include "future revolutionary new capabilities for DoD," as well as training students "for the defense workforce" to enable "long-term relationships between university researchers and the DoD". In short, while it is asserted that proposals for basic research are welcome, it also becomes clear that successful proposals will articulate outcomes relevant to defense interests.

Basic research is also often discussed with explicit reference to battlefield and military applications. In the same Vannevar Bush fellowship cited above, scientific areas set out in the solicitation include numerous examples of the language of warfighting, for example "Fundamental research in neural activity […] can also lead to the development of brain-machine interfaces (BMIs) to facilitate the integration of the warfighters and future AI agents." (p.12) In another example, a funding call from the Army Research Office, for its Historically Black Colleges and Universities program, calls for research on Lidar Remote Sensing, which the call says "could be used" for "remote sensing over the battlefield, e.g., in support of field artillery targeting" (#3). Finally, another "basic research" solicitation promises funding for the so-called "Internet of Battlefield Things," to enable "dynamics that exceed human op-tempos" the "survivability of mission assets" and handling the exigencies of an "IoBT-enabled battlespace" (#4). This solicitation notes that while the "core" research is to be funded as basic research, there is also a path to funding for applied research as an "Enhanced Program". In sum, grant solicitations within the corpus framed as funding for basic research at the same time steer research directions towards military applications, while building relations between university-based researchers and the DoD.

## How the basic/applied distinction is obviated

Finally, there are many programs funded as either Basic or Applied research, with little apparent distinction made between the two. A Navy funding call states that "ONR is interested in receiving white papers and proposals in support of advancing artificial intelligence for future naval applications. Work under this program will consist of basic and applied research…", including "AI enabled training environments that test warfighter skill" (#5) Many other solicitations invite proposals addressed explicitly to military applications; for example a funding call for "Air Delivered Effects" from the Air Force Research Laboratory's Munitions Directorate includes a research funding area on "Autonomous Target Recognition." Subgoals include "Approaches for training AI/ML or traditional algorithms with synthetic target data that result in good target recognition performance when using real target data", and "autonomous handoff of target cue information from intelligence, surveillance, and reconnaissance (ISR) or fire control sensors to weapon[s]" (#6). The Army Corps of Engineers asks for better "sensor payloads" for "drone swarms" to "monitor human […movement…]" (#7).

---

[9]This format refers to citations in the grants dataset. We provide all cited documents in the supplementary data, along with a manifest that links the citation index to the relevant document and page number (if applicable).

We look in more depth at calls for research directed towards military applications below, in our examination of Defense Advanced Research Projects Agency (DARPA) calls for research in AI/ML. Here the point is to underscore the extent to which DoD grant solicitations, including those framed as basic research, are saturated with reminders of the department's aim of expanding the warfighting capacities of the U.S. military. Research characterized within university laboratories as basic, relevant to a wide and possibly indeterminate range of applications, is effectively steered towards military ends through indications that proposals will be ranked in terms of their relevance to DoD priorities. At the same time, DoD investments extend beyond funding for particular research projects, to building out and securing the network of relations, and financial dependencies, that hold the military-academic complex in place. In the following section we look at a series of grant solicitations over time, to trace other devices through which cycles of problem formulation and promised technological solution are sustained.

## From problem to solution, or solution to problem?

Our dataset spans 16 years, allowing us to track trends in AI research topics across this period. To ground our analysis, we examine grant solicitations in three programs, each with a different disciplinary focus, all explicitly categorized as "basic research". Firstly, the Multidisciplinary University Research Initiative (MURI) solicits research yearly on "high priority topics and opportunities,"[10] primarily in STEM fields. Secondly, solicitations for the Vannevar Bush Faculty Fellowship (VBFF) fund faculty fellows each year, typically on the same topics year-to-year ranging from artificial intelligence to materials science to quantum computing (though with some major changes discussed below). Finally, the Minerva Research Initiative is devoted to "supporting social science for a safer world", focusing on topics of "strategic importance to the U.S. national security policy."[11]  The fact that these three programs do not focus solely on AI or even computer science, but instead reflect broad areas of military research interest, allows us to examine AI's position in the military imagination, its relative importance among other fields of study, and its impact as a method *within* other fields and relative to outside trends in academic and industrial AI research.[12] Notably, we see how, especially after 2021, AI increasingly becomes entangled with other fields, where it is seen either as a necessary part of those areas of research, or even their ultimate goal.

---

[10] https://www.onr.navy.mil/education-outreach/sponsored-research/university-research-initiatives/muri

[11] https://minerva.defense.gov

[12] We also found support for the idea that these grants are representative of DoD priorities. In slide 10 of a presentation outlining the scope of DoD basic research (#8) from the  director of the DoD Basic Research Office, the three initiatives are the first listed under the category of "Research Funding to University Laboratories"

Much of these solicitations takes the form one might expect for basic research, setting out aspirational capabilities for computational systems to demonstrate. One of the most common frames through which AI research is solicited across grant series is with reference to biology, where biological capabilities are often taken as very literally comparable to computational approaches: one solicitation from the VBFF series suggests that further work in neuroscience might help us "make giant steps towards general AI". (2023, #9) As others have argued, biology naturalizes and legitimizes the horizon of what AI might one day achieve.[13] If primate vision can already overcome "the inherent tradeoff between area of coverage and degree of resolution," this is a capability "worth emulating by directing foveas […] to salient points in a visual scene" while still being able to "process the contextual content in the periphery" (MURI, 2009, #10). If biological organisms possess a desired capability, in other words, only a supposed lack of basic research stands in the way of computational approaches capable of doing the same.

The analogy between computation and biology is also worked in the reverse direction, attributing desirable computational properties to biological subjects: cells are seen to have "a mastery of low-energy information processing" (MURI 2017, #11), primate cognition is seen as being "robust to adversarial attacks" (MURI 2022, #12) and octopuses exhibit "efficient learning" in contrast to "brute force" machine learning approaches (MURI 2019, #13). In ascribing these properties to biological systems, the writers of these solicitations are able argue that the capabilities they desire already exist and should be translatable to computational systems, while ignoring the possibility that functionality inspired by biological systems may be fundamentally incompatible with computational methods.

## One small problem

Solicitations that outline an aspirational capability employ a recurring rhetorical move that we call the "one small problem" caveat. Here, affirmation of the promise of an existing approach is followed by assertions of its limited scope, and the need for research that would allow the approach to work under more "realistic" conditions. For example, when discussing approaches based on game theory, solicitations assert that it models how "emotionless geniuses" behave, rather than "average people with emotions" (MURI 2011, #14), and assumes a "stable world/environment", which "may not be guaranteed" (MURI 2015, #15). Similarly, reinforcement learning, despite its "dramatic success" at games like chess or Go, is of uncertain applicability to "problems of practical interest." (MURI 2021, #16). This framing balances two competing incentives: emphasizing certain methods' promise and past successes and thus presenting them as viable directions for future work, while at the same time directing that future work towards military contexts in which their "one small problem" requires solution.

Over time in these series, the "one small problem" caveat increasingly narrows its focus to machine learning approaches within AI. In 2011, MURI topic 22 solicits work on game theory with the open-ended goal to "handle imperfect information" by "modeling deception." (#14) By contrast, later solicitations assume much more specific approaches in vogue in the wider research landscape at the time: one solicitation in 2017 refers to "semantic energy" , as well as "deep learning and ensemble modeling approaches" for characterizing graphs (#11); another from 2024 focuses entirely on generative adversarial networks (#17), with each technique appearing in solicitations shortly after their prominence in academic discourse. We see residual effects in the social science focused Minerva grants as well: in successive revisions of their 2013 solicitation(#20), "belief formation" changes to "belief propagation", the name of an algorithm used in the then-popular kind of graphical or Bayesian models. Through these framings, outstanding problems in the application of computational methods to DoD interests can be formulated in terms that affirm the value of previous investments while setting out new directions for further research, updating approaches to problems in ways that align with the changing interests of the research community.

---

[13] The literature here in the history and sociology of science is extensive; see for example Haraway 1989, Helmreich 1998, Dhaliwal et al 2023.

## AI as a durable frame

This narrowing of focus to specific techniques reflects the increasing prominence of AI, as even the conception of problems is set out in the terms, methods, and affordances of AI and machine learning. AI then becomes a *durable frame*: a way of conceptualizing problems over the span of multiple years, and a precedent that any new approach should reference.

In the MURI series, the most notable early example of a durable frame is control theory, prominent in the late 2000s, which faded as big data and network science took over around 2012. Network science remains a durable frame in the VBFF dataset as AI emerges: the 2021 rewrite of topics in the VBFF series (#19) elevates "Networks and Artificial Intelligence" to the first topic of concern, and artificial intelligence is discussed as a field that can be revolutionized using network science. In the Minerva series, quantitative methods can be seen as a durable frame throughout the series, as solicitations for better quantitative methods occur at the top of the list of topics in many years. Quantitative methods are held up as being "more consistent and reproducible" than qualitative methods, offering "well-defined comparisons" (2013, #20). This holds even as the authors of later Minerva solicitations acknowledge that "quantitative" does not mean "objective," that changes in quantification and metrics result in different insights, and that quantitative work may not be "grounded and ethnographically representative" (2022 #21).

By the end of the series, AI is the most prominent durable frame in this dataset. In the 2024 VBFF solicitations (#1), we see invocations of AI in multiple focus areas — applied mathematics, neuroscience, and materials science — which all mention AI's importance in their fields. In the neuroscience topic, we see a shift in the focus of solicited research over time toward AI. In 2016, the first year for which we have descriptions of the VBFF topics, the neuroscience area focuses on supporting "cognitive effectiveness and emotional resilience" for warfighters in the face of sleep deprivation and the "growing volume of data presented by today's weapon systems," along with the development of brain-machine interfaces to assist with recovery from traumatic brain injuries (#22). In subsequent years, however, these focuses shift—while the desire to support warfighters with information overload remains, the rest of the topic is dedicated to prospective applications to AI: research into the brain and its function are now for inspiring new concepts in AI, and brain-machine interfaces are now meant to integrate warfighters with AI systems, rather than support their recovery from injury.

## Inevitability amid ambivalence

Durable frames, in their conception, require successes to point at in order to justify their continued funding and interest. The successes attributed to AI in the series often rely on discourse familiar in more consumer-facing applications, including the inevitability of the revolution that it brings. AI's inevitability, however, is based in claims of its progress" that are, at best, only vaguely connected to the solicitation's intended field of application. One MURI topic in 2020 (#23), "Adaptive and Adversarial Machine Learning," rests its assertions that machine learning "improves overall effectiveness" on the "success of autonomous cars on US roads," despite the fact that no autonomous cars in 2020 had moved beyond trial programs in narrowly defined environments (Piper 2020). Further, as the solicitation notes, deep learning can be "easily defeated by minute perturbations" in operating conditions. Similarly, another MURI topic in the following year (#16) states baldly that "in 10 years, many cyber tasks will be carried out by autonomous systems," without citing any evidence or source for this prediction, while noting that deep learning provides few behavioral guarantees, and key technical assumptions are "commonly […] violated in the real world." In these cases, the authors of the solicitations appear firmly convinced that AI and deep learning are inevitable, despite their own articulation of the (potentially fundamental) flaws with these approaches.

The juxtaposition of inevitability with the acknowledgement of outstanding problems represents a significant and recent rehabilitation of AI's image in the military imagination. Until about 2020, the technology undergirding the modern conception of AI, deep learning, was considered wholly unsuitable for military problems. Solicitations discussing deep learning in 2017 and 2018, rather than posit it as promising using "the one small problem" caveat, instead point at its "crippling fragility", and imply that deep learning with its large data requirements is not "physically viable" for controlling military systems (MURI 2018, #24). Several solicitations go so far as to forbid deep learning approaches for successful submissions. Multiple Minerva solicitations also express an implicit uncertainty about the efficacy of AI solutions, while nonetheless affirming their inevitability in statements like: "regardless of whether possessing AI capabilities equates with power, the world finds itself in a race to develop and deploy these technologies" (2020, #25).

In its growing prominence in the framing of DoD grant solicitations, AI demonstrates a departure from previous forms. Despite the lack of clear (public) successes in the military domain, it is still increasingly seen as inevitable by multiple communities, who at the same time demonstrate their awareness of the fundamental, deeply rooted pitfalls in adopting it: the lack of behavioral guarantees and "crippling fragility" of deep learning systems, the potential for ungrounded quantitative analyses that bear the sheen of objectivity, and the massive costs of data and computational resources. These problems become acutely consequential when we move from domains of primarily laboratory-based research to the arena of warfighting.

# Constructing the (AI) battlefield

Along with the positioning of DoD-sponsored programs in relation to topics of interest framed as basic research, grant solicitations include explicit calls for imagined applications of artificial intelligence technologies to the battlefield. In this section we offer a reading of documents selected from the results of a Segmented Document Search on the keyword 'AI,' filtered for the agency name DARPA. The 101 results returned by the search include Broad Agency Announcements issued from March 2010 to June 2023.

The Calls discussed below include direct reference to AI or ML, as well as explicit indications of anticipated battlefield use. We approached our reading of these documents with three overlapping questions in mind. Firstly, what "small problems" in existing warfighting capacities do the calls reference in order to motivate further research and development in AI/ML? Secondly, how do the calls frame the future of warfighting, whether as necessary responses to the operations of others or as aspirational optimizations of US military strategy? And finally, how are these problems and their imagined solutions located in histories of modern warfighting and its contemporary formations? We address these questions under three themes that mark both longstanding and emergent areas of military interest: Situational Awareness, Human-Machine Teaming, and Distributed Warfighting/Lethality.

## Situational awareness

Continually confounded by the "fog of war" (Shapiro 2005), situational awareness (SA) is a foundational concern of military doctrine. SA is defined by Micah Endsley, former Chief Scientist of the United States Air Force, as "the perception of environmental elements and events with respect to time or space, the comprehension of their meaning, and the projection of their future states" (1988; for a critique see Suchman 2023). Military doctrine takes information gathering, formalized as Intelligence, Surveillance, and Reconnaissance (ISR), as key to enabling situational awareness through network infrastructures and targeted surveillance. Consistent with the automation of computation in the mid-twentieth century, these capabilities are conceptualized as recursive procedures that follow the hierarchical ordering of command and control. This runs from the formulation of strategy by senior officers, to the tactical operations of front-line commanders, through to the perceptions and actions of individual warfighters. In the contemporary moment, our corpus shows how optimization of situational awareness is to be achieved through increasing reliance on the application of AI (principally ML) in ISR.

In March 2010, DARPA issues a Broad Agency Announcement (BAA) for the "Mind's Eye Program," with the aspiration to develop machine capabilities in "visual intelligence." (#26) The announcement opens with a review of the limits of the current state of computer vision, in its reliance on statistical computations based on benchmark datasets rather than models of perception and cognition in realistic environments. As in the 2009 MURI call discussed above (#10), humanlike visual intelligence is characterized as the capacity to represent observed events in "the mind's eye," in order "to manipulate those imagined scenes mentally to solve problems." The (re)turn to mental modeling indexes the 'one small problem' of the limits of ML-based techniques for computer vision, and accompanies a shift in focus from object to action recognition. In this case advances in computer vision are aimed at the creation of a "smart camera," to be deployed as a "payload" on Unmanned Ground Vehicles (and potentially other platforms), with sufficient "machine-based visual intelligence" that it can report on activity in an area of observation. The aim of the research program is to develop the "perceptual and cognitive software" that would equip the camera with "the capabilities necessary to perform surveillance in operational missions." The aspiration, in other words, is to encode the specifications for situational awareness into the camera itself, a move that requires the translation of detectable signals into warfighting's normative frames.

The BAA goes on to list the requisite capabilities in terms that anticipate their translation to code, beginning with the "recognition of primitive actions that take place between objects in the visual input, with a particular emphasis on actions that are relevant in typical operational scenarios (e.g., vehicle APPROACHES checkpoint; person EXITS building)." The system should learn "generally-applicable, invariant spatiotemporal patterns that make up these actions directly from visual inputs based on few labeled examples and with minimal supervised learning," and be able to apply the patterns learned in one scene to another. Those capacities should be iteratively refined through continued exposure to visual inputs during the camera's operation, such that the system "develops a sense for what is ordinary and out-of-the-ordinary through continued exposure to a particular visual scene." In addition to these imagined ML-based capabilities, the system should incorporate "reasoning" functions that integrate "stored and acquired knowledge of visual objects as well as symbolic knowledge" to enable it to issue "alerts to activities of interest, as defined by concepts the device is tasked to detect." Examples of such concepts include the introduction of an object into the scene, or the hand-off of an object from one person to another. The system should be "taskable" through simple text commands and provide alerts via text messages, as well as able to "explain its reasoning by displaying relevant video segments for what has been observed." Finally, and most ambitiously, the system should perform interpolation (inferring actions that are not directly observed, e.g. an object carried into a building that is absent on the person's exit from the building must have been left inside), and prediction (projecting next actions based on previously observable cases, e.g. a boy carrying an object and approaching a soldier may be expected to hand the object over) (see Figure 1).[14]

---

[14] While presented as a straightforward exchange in Figure 1, we could consider this encounter in the context of contemporary operations (notably current IDF operations in the Occupied Palestinian Territories) where, the civilian status of children under international humanitarian law notwithstanding, the child's approach might result in the use of lethal force on the part of the soldier.

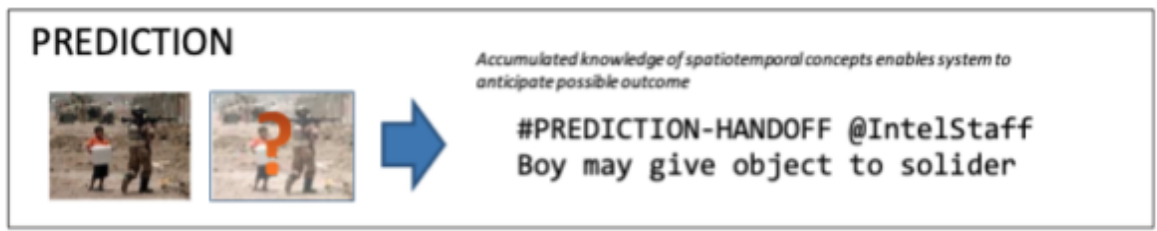

**Figure 1**: A diagram from a grant proposal demonstrating the prediction that a boy approaching a soldier may hand him an object, based on previously observed cases.

The aspiration to advance from object to activity detection begs the question of to what extent even object recognition in the context of warfighting is a solved problem, particularly when the objects in question are not specifically military (e.g. armored tanks) but ordinary civilian objects (e.g. in this case a box). But we can see here a desire to expand the range of algorithmic processing to encompass surveillance operations aimed at the detection of anomalies in so-called 'patterns of life'.[15] In its explicit mandate for the inclusion of "reasoning" functions, the call anticipates a niche but persistent thread within the AI research and development community away from a reliance exclusively on statistically based ML toward a 'hybrid' approach that restores techniques of knowledge representation and symbolic logic marginalized with the rise of data-driven approaches (Marcus 2020).

Over a decade later, the vision of algorithmic analysis of ISR data as a hybrid of ML and symbolic reasoning is reiterated in a September 2022 BAA from DARPA's Information Innovation Office (I2O) (#27). The call for "Environment-driven conceptual learning (ECOLE)" sets out an ambitious program for situational awareness, object recognition, activity-based intelligence, and "pattern of life" analysis. Now, however, the human analyst is restored to the frame. The program aims to "enable human-machine collaborative analysis of image, video, and multimedia documents during time-sensitive, mission-critical DoD analytic tasks [so that the machine can] readily learn a new symbolic representation through interaction with its human partner." The move to "human-machine collaborative analysis" could be read as well as an acknowledgement of the limits of the hybrid approach imagined in 2010, as the 2022 BAA calls for "an extensible framework capable of instantiating an arbitrarily large (i.e. global-scale) web of knowledge representations." The hope for a comprehensive encyclopedia of relevant information is an aspiration that dates back to the mid 1980s and the Cyc project, a four decade-long effort to codify the "common-sense knowledge" assumed to be the foundation of human reasoning (Fisher 2024). Here, the call attributes the 'small problem' of training in ML and the continued need for human feedback to the limits of knowledge representation (a problem that data-driven statistical approaches promised to resolve). But while initially imagined as a replacement for human experts, knowledge representation is reconceived here as a form of human-assisted machine learning, which brings us to our second theme.

---

[15] A growing body of critical scholarship and documentation of tragic killings challenge the premise that 'patterns of life' comprise a reliable basis for targeting; see for example Suchman et al 2017, Wilke 2017.

## Human-machine teaming

Along with the aim of greater machine autonomy, contemporary investment in AI is justified within the research and development community as a step toward "harmonious partnership between humans and AI."[16] Embraced by the DoD, the aspiration to what is named "human-machine teaming" is a continuation of a cybernetic imaginary that positions the human as the "machine in the middle" of complex military systems.[17]

In March of 2019, a BAA from DARPA's Defense Sciences Office calls for proposals on the "Science of Artificial Intelligence and Learning for Open-world Novelty (SAIL-ON)." (#28) The problem to be solved is formulated as the limited progress of AI systems in "understanding the most important component of the environments in which they operate: humans." The BAA returns to a paradigm of mental models as the basis for successful interaction.[18] Among other deliverables, the BAA asks for simulations of "agents" that demonstrate the requisite social skills, implemented in a virtual testbed with input from standardized "sensing channels" and output through "communication/action channels" made available to human teammates through standardized interfaces. The testbed should demonstrate agents' capacities to "operate in increasingly complex and specialized environments, be adaptable to sudden perturbations in the mission or team (like the loss of communication with a key teammate), and use noisy multi-channel observations to represent the world and do complex inference and prediction." By confining the demonstration to a simulation, rendering contingency as complexity, specialization, perturbations, and noise (all terms with familiar computational translations), and defining social skills as the alignment of mental models, the call for "open-world novelty" remains within the arguably closed-world of information processing psychology.

Modeling of mental states is expanded in a November 2020 DARPA call for Post Doctoral Fellowships in the domain of Emotion Recognition, based on analysis of "ego-centric video captured from head-mounted cameras."(#29) The premise is that "estimating and modeling user emotions is important for effective human-computer teaming." More specifically, the aspirational capability involves "new techniques to determine the emotional state of a user of a head-mounted camera based on the ego-centric video, audio, and inertial data it collects […] The goal of this project is to develop new techniques that integrate any and all cues to a user's emotional state into a continuous, real-time estimate of user affect." Applicants are reminded that this "ego-centric video data" does not include the user's face, excluding the usual source data for so-called "emotion recognition". The research challenge is to move beyond affect or sentiment analysis based on facial images to inferences from first-person perspective images of a scene, recorded speech, and head motion, gesture, and body position, as well as "how a user's emotional state may be reflected in the emotional reactions of other actors in the scene, actors whose faces are visible from the user's perspective." Despite the fact that the validity of emotion recognition based on facial images is highly questionable,[19] the call proposes to extrapolate further from that approach, using inferences attributed to the user as a basis for their own, not directly observable, affective state.

---

[16] HAI 5 Agenda Outline https://hai.stanford.edu/hai-5-agenda-outline

[17] As Edwards (1996: 179) observes, the rise of cognitive psychology during the latter half of the twentieth century "reconstructed both humans and animals as cybernetic machines and digital computers." Edwards shows how, within this imaginary, humans become incorporated into systems of command and control as enabling components.

[18] For a critique of this premise see Suchman 2007, chapter 7.

Making the interior states of warfighters available for assessment and management is matched by a concern for the "introspective" capacities of machines. To realize the promise that weapon systems can become collaborative and trustworthy warfighting partners, DoD calls reiterate the need for technologies with the capacity to adjust and fine-tune their own operations in response to unfolding circumstances. A BAA released in August 2021 by DARPA's Information Innovation Office, titled "Learning Introspective Control," "aims to develop machine learning-based introspection technologies that enable physical systems, with specific interest in ground vehicles, ships, drone swarms, and robotic systems, to respond to events not predicted at design time." (#30)  In May 2022, an Information Innovation Office BAA on "Assured Neuro-Symbolic Learning and Reasoning" reiterates 'one small problem' within the capabilities of existing warfighting systems, namely that "high levels of autonomy remain elusive" due to the limits of ML-based approaches (#31). This call responds to the Defense Science Board Report on Autonomy (Office of the Undersecretary of Defense 2016), emphasizing the importance of trustworthiness: "Informally, trust is an expression of confidence in an autonomous system's ability to perform an underspecified task [and is] key to DoD's success in adoption of autonomy." The call is focused on the question of autonomous systems to be used in ISR, logistics, planning and Command and Control, where "The purported benefits are many, including – (1) improved operational tempo and mission speeds; (2) reduced cognitive demands on [the] warfighter in operation and supervision of autonomous systems; and (3) increased standoff for improved warfighter safety," that is enhanced action at a distance. Machine autonomy, in sum, is a requirement for human-machine teaming, which in turn enables and legitimizes the further automation of weapon systems.

In the context of the international debate over the legality of automated targeting, and civil society campaigns against autonomous weapon systems, the invocation of "human-machine teaming" serves as an assurance that human control will govern the incorporation of AI technologies into the kill chain. Critics have argued, however, that there is an irresolvable contradiction between increasing the speed of warfighting through automated targeting and adherence to the requirements of International Humanitarian Law.[20]

---

[19] For a critique of computational approaches to "emotion recognition," see Stark and Hoey 2021.

[20] See for example Holland Michel, 2023.

## Distributed warfighting/lethality

In the years following the terrorist attacks of September 11, 2001 the focus of US military strategy shifted from superpower conflict to what critical scholars have designated as "everywhere war" and the increasing use of Special Operations Forces (Chamayou 2014; Masco 2014). The turn to Special Operations and a program of continuous, distributed counterinsurgency is accompanied now by the revival of competition for military dominance, framed as a new arms race between the US and China. Both developments support a shift from a traditional focus on maneuvers at "the front" to operations conducted "on the edge," which in turn require new modes of "command and control" across forces. Conjoined with the aspiration to greater distribution of command and control is a growing concern for integration across the longstanding divisions between the various armed forces, intensified by the increasing expansion of all forces into multiple 'domains' of warfighting (air, land, sea, space, and cyber), imagined in the vision of Joint All Domain Command and Control (JADC2) as a set of interlocking and interoperable 'systems' (U.S.Department of Defense 2022a).

Consonant with these developments, a June 2017 BAA from DARPA's Tactical Technology Office (TTO) titled 'Innovative Systems for Military Missions' (#32) calls for demonstrations and research programs aimed at "cross domain systems" that "validate revolutionary precision engagement capabilities." The requirements for precision engagement include "the development of persistent, global surveillance architectures; real-time data updates, at scale; provision of real-time, decision quality information; and the demonstration of novel approaches that support rapid and affordable integration," with a focus on the deployment of "swarm technologies." Two years later, in March 2019, the aspiration to command and control "at the edge" is evident in a BAA issued by DARPA's Microsystems Technology Office on 'Real Time Machine Learning,' addressing the problem of how ML systems can incorporate "new datasets in the field." (#33) A joint effort with the National Science Foundation (NSF), the aim is the "rapid development of energy efficient hardware and ML architectures that can learn from a continuous stream of new data in real time."

In a February 2019 BAA, DARPA's Strategic Technology Office (STO) goes directly to the point, stating that the STO "is seeking innovative ideas and disruptive technologies that provide the U.S. military increased lethality in an era of eroding dominance."[21] (#34) The announcement goes on to explain that U.S. dominance in the development of sophisticated weapon systems is now being "challenged by peer competitors." The response must be a "new paradigm that values 'lethality' over monolithic system dominance. Whereas dominance is measured by comparing capabilities across systems, lethality is measured by the ability to deliver a desired effect at will, regardless of the system or systems of systems involved." This lethality is to be delivered through 'Mosaic Warfare', enabling "fast, scalable, adaptive joint multi-domain lethality. It is the disaggregation of effects chain functions (e.g., Find, Fix, Target, Track, Engage, and Assess or F2T2EA) across a heterogeneous mix of manned and unmanned platforms from all domains." A BAA in October 2021 from the STO on Mosaic Warfare continues the program (#35), with an emphasis on the achievement of a "disaggregation of capabilities and re-composition of adaptable effects chains at mission speed." In military doctrine the most prominent "effects chain" is a successful "kill chain," a sequence of intelligence analyses and decisions leading up to a potential strike, in a process promoted as an appropriate application of AI-enabled optimization.[22]

---

[21] The premise that US military dominance is "eroding" is challenged by a comparison of military spending internationally. According to the Stockholm International Peace Research Institute, the 2023 US defense budget exceeded that of the 10 next most heavily armed countries in the world combined (including both China and Russia). See https://www.pgpf.org/chart-archive/0053_defense-comparison

We see, in sum, the emergence of a paradigm of US military dominance across two, parallel constructions of the geopolitics of warfighting, one characterized by ongoing, "low level" but no less lethal, operations of so-called counter-terrorism and counterinsurgency against "irregular" and often non-state actors, the other by a return to superpower conflict, most imminently in the Indo-Pacific region and focused on China. Within the DoD's research and development agenda there is a place for AI-enabled weapon systems in both, and both provide justificatory grounds for research directions well aligned with the current interests of the commercial and academic AI research community.[23]

## Conclusion: AI research and militarism

Our examination of a selection of documents from the U.S. Department of Defense grant solicitation corpus is aimed at tracing how requests for proposed research help to secure the institutional relations that form the military-academic complex. Our characterization of those relations as a form of mutual enlistment is meant to emphasize the two-way traffic through which technopolitical imaginaries are formed out of shared interests in the promotion of research agendas that, however loosely coupled, are positioned as serving U.S. national security interests. We have argued that rather than delineating two separable arenas of research and development, the distinction of "basic" and "applied" effects a trading zone in which research and development can be steered in the direction of military imaginaries of future warfighting, which in turn are shaped by promises advanced within academic and commercial research and development networks. The trope of basic research in this context offers shelter from engagement with moral questions raised by military applications, while obscuring the connections that implicate researchers in the wider complex of U.S. militarism. A time series analysis of programs framed as basic research illustrates the rhetorical moves through which the results of previous investment are affirmed, at the same time that new problems requiring further investment are identified. And, an indicative analysis of a subset of DARPA calls focused specifically on AI's promise for the battlefield highlights DoD imaginaries of technological solutions to longstanding problems, refigured and arguably exacerbated by previous investments in data-driven warfighting.

---

[22] On the evidence for the application of this logic by the Israeli Defense Force and its consequences in Gaza see Abraham 2024; Hables Gray 2024; Suchman 2024.

[23] See Congressional Research Service 2021. Current DoD-funded projects include the Replicator Initiative, aiming to develop "swarms" of autonomous combat drones deploying AI for the identification and targeting of "threats," an Air Force project to develop a fleet of 1,000 AI-enabled fighter jets (Hicks 2023), and Project Maven, an AI-enabled, computer vision system for the automated analysis of surveillance data in service of threat detection and target recognition, now a central project for the commercial military contractor Palantir (see Harper 2024).

Within the US Department of Defense and wider national security establishment the premise that "[t]he safety and prosperity of our democracy depends on our ability to innovate and stay ahead of the technological advances of our adversaries" is taken as incontrovertible.[24] Held in place by doctrines of deterrence-based defense, this premise justifies the continued expansion of the "Department of Defense, the defense industrial base and the array of private sector and academic enterprises that create and sharpen the Joint Force's technological edge" (U.S. Department of Defense 2022b). As Pentagon expert William Hartung (2024) observes, taken together this comprises "a military conquest (so to speak) of America's research and security agendas." Recent media coverage of the incorporation of computer vision systems into the control of armed drones in Ukraine, and of ML systems into target generation in Gaza,[25] along with pronouncements about the threat of China's investments in AI,[26] reinforce the premise that an AI arms race is inevitable and that the acceleration of AI-enabled warfighting must be a DoD priority. Alex Karp, CEO of military contractor Palantir (and clearly an interested actor), has declared that AI-enabled warfare and autonomous weapons systems have reached their "Oppenheimer moment."[27] It is perhaps appropriate, then, to remember the legacy of nuclear weapons development, including the emergence of scientific fora dedicated to nuclear arms control and disarmament (Hartung 2024).

Our aim in this analysis is less to attribute intent to actors within the DoD's research funding agencies or in the academy, than to trace the circuits through which obstacles to military dominance are formulated as problems for which the AI research community can offer solutions. Academic research is considered an integral part of the "DoD Innovation Ecosystem," currently itself the target of highly publicized initiatives to re-engineer its operations in ways aimed at facilitating returns on DoD R&D investments in data-driven, AI-enabled warfighting.[28] Grant solicitations are a key site for the articulation of intersecting interests that not only inform, but are in important respects constituted through, resulting networks and relations. This mutual enlistment is crucial to the perpetuation of the military-industrial-commercial-academic complex, and to the technopolitical imaginaries of security through military domination that keep public funds flowing to projects in more efficient killing and destruction, and away from experiments in creative diplomacy, de-escalation, and alternatives to militarisation as a basis for our collective security.

---

[24] Shyu and LePlante 2024.

[25] See McKernan and Davies (2024).

[26] This assessment is contested by researchers who argue that the significance of China's technological progress has been overstated in the interest of promoting an AI arms race. See Toner, H., Xiao, J., & Ding, J. (2023), Hartung (2023).

[27] Karp, A. and Zamiska, N. (2024) Palantir is the primary contractor for Maven Smart Systems, the extension of Project Maven.

[28] On the growing role of venture capital and Big Tech in DoD procurement restructuring and their influence on AI research, including in the academy, see Whittaker 2021.

# Acknowledgements

We are grateful to Dan Bateyko for collecting the original dataset and supporting this project. We are grateful to Carnegie Mellon University for their enthusiastic institutional support for the military, which in large part inspired this paper. Widder gratefully acknowledges the Digital Life Initiative at Cornell Tech. Gururaja thanks Emma Strubell, without whose support he could not have participated in this work, and appreciates the military funding that partially supports his work, though it was not specifically associated with this project.